# What Do Neural Nets and Quantum Theory Tell Us About Mind and Reality?

Paul J. Werbos
National Science Foundation*, Room 675
Arlington, Virginia, USA 22230
Pwerbos@nsf.gov

## 1. Introduction

The organizer of this conference, Dr. Kunio Yasue, invited people from many disciplines to address certain basic questions which cut across these disciplines: "How can we develop a true science of consciousness? What is Mind?" This paper was invited to the session on quantum foundations, which was also asked to address: "What is Reality?"

The literature on consciousness contains many discussions about what we can learn from modern neural network theory and quantum theory, in trying to answer these questions. However, those discussions do not always account for the most recent insights and developments in those fields. Even those authors who deeply understand all the relevant disciplines would find it difficult to write a paper which is intelligible to people in other disciplines, but also does justice to the real technical details.

Because of this communications problem, I will write this paper in a relatively informal way. The bulk of the paper will be an edited transcript of the talk which I gave at the conference, with references added to provide at least some technical support. Section 3 will contain new thoughts, stimulated in part by discussions at the Quantum Mind conference in Arizona, later in 1999. The views expressed here are only my views, not the official views of NSF or of the US government.

## 2. Transcript of Talk

In his introduction, Dr. Yasue mentioned that Paul Werbos is a Program Director at the National Science Foundation, the primary agency of the US government for funding basic research across all disciplines. He studied physics under Dr. Julian Schwinger, winner of the Nobel Prize for quantum electrodynamics along with Feynman and Tomonaga. He is best known for the original discovery of backpropagation, the most widely used algorithm in the field of artificial neural networks.

Thank you, Kunio. I am very grateful to have a chance to speak to you here in Japan.

Before I begin, I must make a couple of apologies. First, I am not really a professional physicist. I *did* have the good luck in graduate school in Harvard to study under Julian Schwinger who, as you say, was the co-inventor of the quantum field theory discussed by many speakers here. In the 1970s, when I studied under Schwinger, many people actually thought he was going crazy, because Schwinger *did not like* the second quantization, the quantum field theory. He felt there must be a better way to do it -- and so, in the 1970s, he worked on a new way of doing quantum mechanics. He called it source theory (Schwinger). He had the framework right at that time, but he did not yet have the details of how to apply it to high-energy physics. So when I was a student they said "This is crazy. The formalism is OK, but it's not practical. It's just metaphysics; don't pay attention to it." But in the last twenty years, I was very happy to find out that this source theory has been developed much further. It is now called the functional integral approach (Zinn-Justin). It is a *third* quantization. It is a whole new way of doing quantum mechanics, and it changes many of the things we have heard here. Quantum field theory today is not what it was twenty years ago.

I have not worked in physics myself since then, but, on my own, I have tried to use my scarce personal time to think of yet another way to do the quantum foundations. I have some wild and crazy ideas for a fourth quantization. I have a few papers on it, but I only have the mathematical framework (Werbos 1989, 1998a, 1999a). I think the framework makes sense, but much work is needed now to develop the practical details. I hope someone here is a physicist interested in working out some of the details, because I

---

- The views herein are those of the author in 1999, not those of NSF, though it was written up on government time.

am not like Schwinger; I will not spend the next twenty years developing the practical details. I would be grateful for any collaborators for the next stage.

But no one pays me to do physics. Actually, I work in the Engineering Directorate at NSF. So here I feel like a humble shoemaker asked to give a talk at the great temple; in one week, I will go back to making shoes -- but the shoes we make are not exactly shoes. We help people develop cars which are cleaner and more efficient, airplanes which are safer and faster, new manufacturing systems, robots, control systems for electric power grids (Werbos 1999b,c). Carefully and slowly we develop real engineering things which must work. That is what they really pay me to do.

So I will begin here by talking about the mind, first. The theme of this conference *is* "consciousness" -- the science of consciousness - and that is what they pay me to do, to worry about intelligent systems and about how this relates to biology. And then I will talk about advanced quantum theory if there is time. I hope there will also be some time to talk about the connection from quantum theory to the brain. Maybe I should say just a few words about that now because I probably will run out of time.

## 2.1. Quantum Theory and the Brain

At NSF, some people want to start a new funding initiative in quantum computing. This is an exciting field. Many people speculate that quantum mechanics can help us do better computing, that we can build a higher level of intelligence if we exploit quantum theory. Many people at this conference have said that with quantum theory, we can explain or produce a higher level of consciousness.

I think this is probably true, but we have not proven it yet. No one has built a quantum system, or designed one which is well-defined, which would really generate such higher-order capabilities. There *are* theoretical concepts for how to use quantum theory to build an associative memory. I think that is what we just heard from Vitiello – some ideas on how to use quantum principles to build or explain associative memory. There is also a person named Grover , who is very famous in quantum computing, famous for *his* design of an algorithm to do associative memory.

But there are two problems here. First, these designs are very theoretical. To create real, working physical systems is much harder than the theoretical physicists used to think. The theoretical physicists tend to work in the second quantization, in a world of pure probability amplitudes. But when you need it to work in real hardware, you need to worry about these horrible quantum thermodynamics issues, which means that you need to think about density matrices. Only recently have people begun to get ideas about hardware which seem to make sense in physical terms. There are ideas, but just beginning. (For some of the recent decisive work on hardware, you may search on names like Gershenfeld, Kimble, Preskill, Lloyd, Wineland and Kwiat on the index at xxx.lanl.gov.)

Second, the more difficult problem is with algorithms. A memory is not a brain. Building an associative memory does not tell us how to build an intelligent system. There is a long distance from knowing how to do an associative memory to proving you can do brains. In many ways, associative memory is much easier than real intelligence. In the questions after that talk someone asked," Are you minimizing energy or are you doing what a brain is doing?" *Of course*, it is not what a brain does! A brain is not a memory. A memory is a useful *part* of a brain but a brain is something much, much bigger.

So now I will try to talk first about my ideas about consciousness and the mind, and then quantum theory, and we will see how far I get.

## 2.2. Consciousness or Mind From a Neural Net Perspective

To begin with, what do we mean by the word "consciousness"? As people have said, there are many, many different definitions of consciousness. In a talk five years ago (Werbos 1997), I tried to discuss six of them:

- o  Consciousness as Awareness
- o Subjective Sense of Existence
- o Consciousness as Intelligence
- o Consciousness Vs. Unconsciousness
- o What About the Soul?
- o Quantum Effects Relevant?

These are just six. I have heard many others at this conference. I do not want to argue about what is the best definition. These are *all* important concepts. Waking and sleeping states – they are very important. But in my talk I only want to talk about one concept. These are all big subjects, so I will focus on *one* question here -- consciousness as intelligence. What is intelligence? What is mind? That is what I want to talk about.

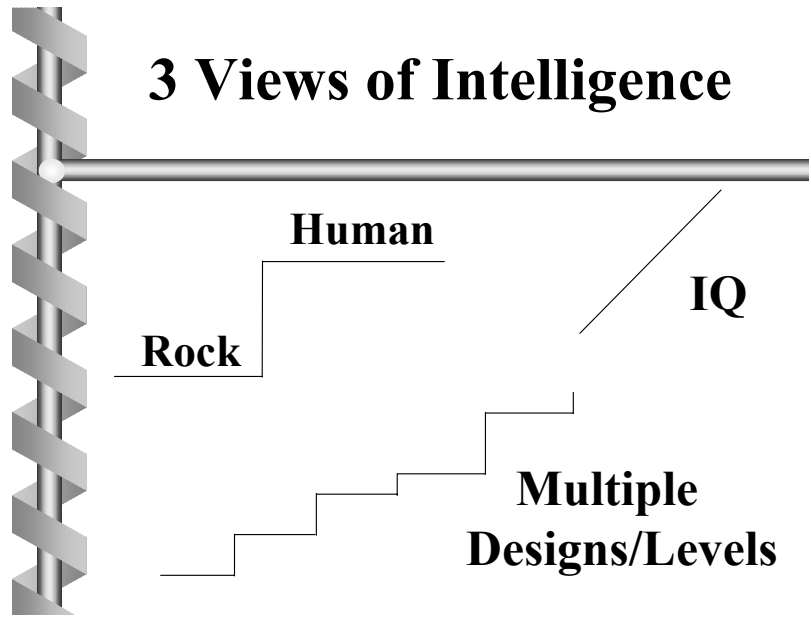

So now let us move to the first slide. If we focus on the idea of consciousness as intelligence, there are still many different points of view to sort out.

There are actually three different concepts or views of intelligence, or of consciousness qua intelligence. The most *common view* I have heard lately is the binary view, illustrated on the upper left of the slide. People look at a computer design ... or they look at a spider... and they ask,"*Is it* conscious? *Or is it* not?" They agree that humans are conscious or intelligent -- I'm not sure why they all agree on that -- but anyway, they all agree on that. They agree that rocks are not intelligent. And then... they worry. "Is this computer system really conscious or *is it not*? A spider – is it really conscious or *is it not*?" This question *assumes* that consciousness is a binary variable, that it is either "yes" or "no." It reminds me of some high school students I knew, when they talked about sex appeal. They said "You have it or you don't." That's it -- it's binary. Well, I'm not so sure it's binary. There might be some matter of degree here.

There is another view, that views consciousness or intelligence as a continuous variable. The stupidest form of this view is the idea of consciousness as IQ -- I don't believe in that, but there are other ways of thinking of intelligence as a continuous phenomenon. Allen Hobson spoke yesterday about *wakeful* consciousness as a *graded* phenomenon. His AIM model presented consciousness as a continuous variable. I am speaking about a different kind of consciousness, but the same principle may apply here. Intelligence may not be binary; it may be graded.

Earlier, David Chalmers talked about panprotopsychism here. Well, there is a very *old* tradition in philosophy called panpsychism. Taoism was like this. They would say that intelligence is present in all things, but in varying degrees. A Taoist would say there is intelligence in the human, the spider, the rock, the tree, the water -- they all have *some* intelligence. It is a question of how much. So I have a funny picture in my mind. I see a philosopher of the West staring at a spider, thinking "Is it conscious or is it not?" And I see an old Taoist master looking at this philosopher and saying: "Is this philosopher conscious or not? Is she aware of what she is looking at? She is looking at a spider. She is not looking at a binary variable." The Taoist would say "Of course there is some feeling in the spider.. but you should be aware of the spider and ask 'What *kind* of consciousness does it have? What is the *nature* of its feeling? What does it feel like? But you should not worry about some binary question in words which make no sense."

There is another group of people who believe in the continuous view -- a much stranger and weirder group, not Taoists, but old-style behaviorists. The old-style behaviorists believed that all animals have essentially the same kind of learning. There was a doctrine which said that... first... intelligence *is* learning. That's a good start. That's not so bad. (Werbos 1994a:3-5, 1994b:682-683.) But then they said... the learning curve is the same for humans, rats, all animals. The humans and rats respond to the same variables; they have the same kind of learning, but the human is a little faster. I think that one reason they believed this was that they could get money to study rats and say that this all applies to the humans. But then they said the same thing for birds and snails, that snails are like humans ... but they are like *slow* humans.

Well... I do not agree with that theory. I think that the right way to think about intelligence -- or about consciousness as intelligence – is what I show on the bottom part of the slide above. I think that intelligence is a *staircase* – a matter of discrete levels for the most part, levels and levels of higher and higher intelligence and consciousness. So we should not ask "Is it conscious or is it not?" We should ask "What is the *level* of consciousness or intelligence?" Now, why do I think this is the right way to think about intelligence? Consider the next slide.

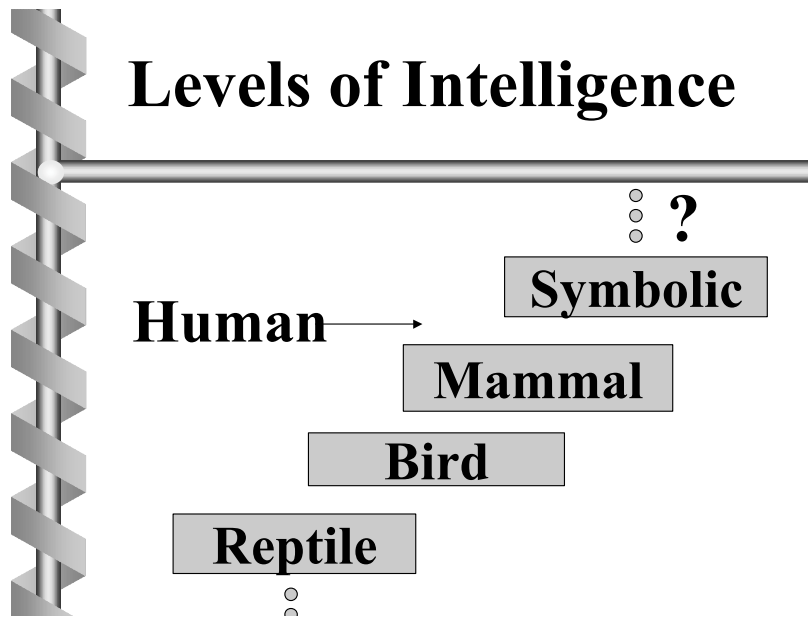

I believe that intelligence is a kind of staircase because this is what we actually see in nature. This is what is real. This is not imaginary philosophy, if you forgive the expression. This is what we really observe in nature. We see reptiles, birds, mammals ... and there is also a kind of intelligence based on symbolic reasoning. That is what built Tokyo -- humans using symbolic reasoning.

Now... I could talk about this slide for a very long time. This is a very important slide. There are many ideas to think about here.

First, I must make some small observation. Some of you may have seen maps of the brain of a rat. You will see that in the cortex of the rat there may be about seven areas for vision. And then you look at a monkey or a cat, and there are more areas. (Arbib: 1025). You may say "Gee, they look very different." But those maps are maps of the neocortex, the highest.. or at least the outermost.. *part* of the human brain. The six-layered cerebral cortex, the neocortex. But the important thing is that *all mammals* have this neocortex. Birds do not have that kind of neocortex. So in a sense all mammals have essentially the same wiring diagram. If you think of *learning* ... if you think of Dr. Matsumoto's "superalgorithm" .. then the basic principles of learning are fairly uniform across the neocortex. Thus in some sense we may say that all mammals are essentially the same. I don't have time to elaborate now.

So now let me talk about strategy. How can we ever build a true "science of consciousness"?

Some people in artificial intelligence (AI) said years ago: "*Real* intelligence is up here, at the symbolic level. So let us try to build an artificial human, by building a machine to do symbolic reasoning." And sometimes they talk about Einstein, and how intelligent he was. "Let's build an artificial Einstein." I think this is the reason why classical AI failed to achieve its highest goals. Classical AI failed to produce true brain-like intelligent systems because they tried to do too much. They tried to go *directly* to the symbolic level, without doing the mammal level first.

There are some people who want us to go *directly* to the quantum/psychic/spiritual remote viewing level. I think that is even worse than trying to go directly to the symbolic level. It is good to *think* about these higher levels, because they are very important... but in order to develop a *science* we need to develop mathematical models and principles that *work*. I think we need to develop the science of the mammal level first, and that will give us the insights we need for better understanding at the symbolic level and even at the levels beyond (the question mark on my slide).

Now if you are a mystic, you may wonder "What can the mammal brain tell us about the deeper human soul?" Well, that is a complex topic. But let me say briefly... there are some mystics who use an expression "As above, so below." Before you can understand the higher levels, they say, you must *firmly* understand the lower level (what is right in front of you), and also understand the analogy between the levels. Thus I claim that the important opportunity, the real opportunity for the science of consciousness today, is to really understand first this mammal brain level, without the soul, the simple basic mammal brain ... that level of intelligence...and to do this mathematically (i.e., to extract the underlying principles, not just the biochemical details) and then see what insights we get regarding the higher levels.
So that is what I have worked for most of my life on, to try to understand the mammal level.

But how can we understand a mammal brain? How can we understand intelligence, at the mammal brain level? Well, I would like to make an analogy, shown on the next slide.

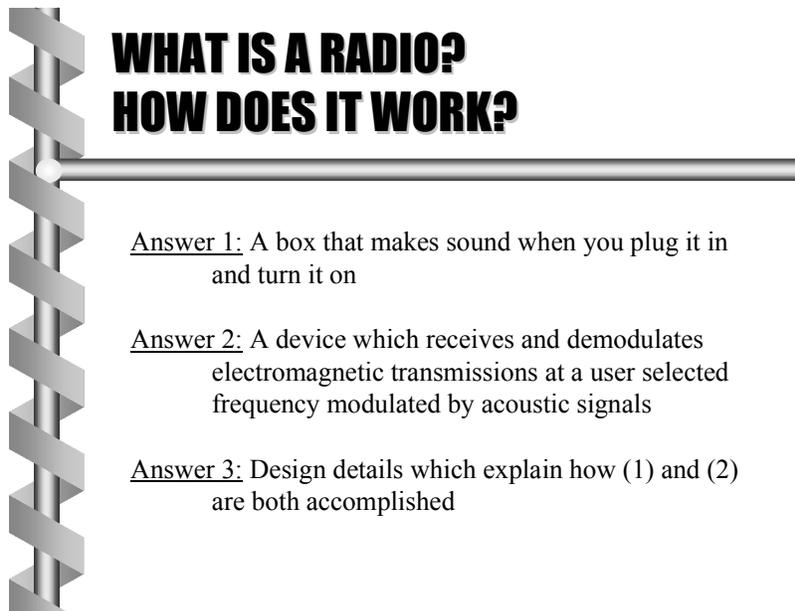

Actually, I am taking this analogy from Charles Gross, a neuroscientist, a student of Karl Pribram's.

In my first course in neuroscience, on the first day, Charles Gross said:" Neuroscience today is like people studying a radio. They buy a thousand radios, to understand how they work. You buy a radio. You turn it on. You pull a tube out... and then the radio whines. You call the tube 'the whine center.' " Then you take a new radio -- throw out the old one into the trash -- it was alive, but you throw it out -- pull out a capacitor, and then you hear a scratch sound. You call the capacitor "the scratch center.'" And then you have a map of the brain where you have the whine center, the scratch center and then you say '"Aha, now I understand the radio.'" But.. you do not really understand the radio.

There are different ways of understanding what a radio is and how it works. There are different ways to answer the question "What is a radio?". At one level of answer, you say "A radio is a box that makes sound when you plug it in and turn it on." This is like the Turing test for consciousness. It is a *descriptive* test. But engineers do not like that kind of definition so much. Then there is what we would call a functional definition: A radio is a device which receives and demodulates electromagnetic transmissions at a user-selected frequency modulated by acoustic signals. I can almost hear some people saying "Isn't that too complicated?" Maybe it is complicated, but this is what a radio *is*, in functional terms. But... for a science... for engineering... we want something even more. We want the design details which explain *how* these characteristics are accomplished, and how they can be replicated... and that is *very* complicated. It *has* to be complicated. I do think it is possible to develop an understanding of consciousness and learning which is simple in the same way that general relativity is simple. Now some people will be very disappointed at a theory which is only as simple as general relativity.... but I think it is very exciting that some of us now see a way to produce such a theory.

By the way, I have one last point to make about this slide. To understand a radio in functional terms, you do not need to know where every screw and bolt is. You don't need *all* of those details. So I'm not talking about knowing every screw and every bolt in the brain.

So now... how can we produce a design-level mathematical understanding of intelligence at the level of the mammal brain, that kind of intelligence? See the next slide.

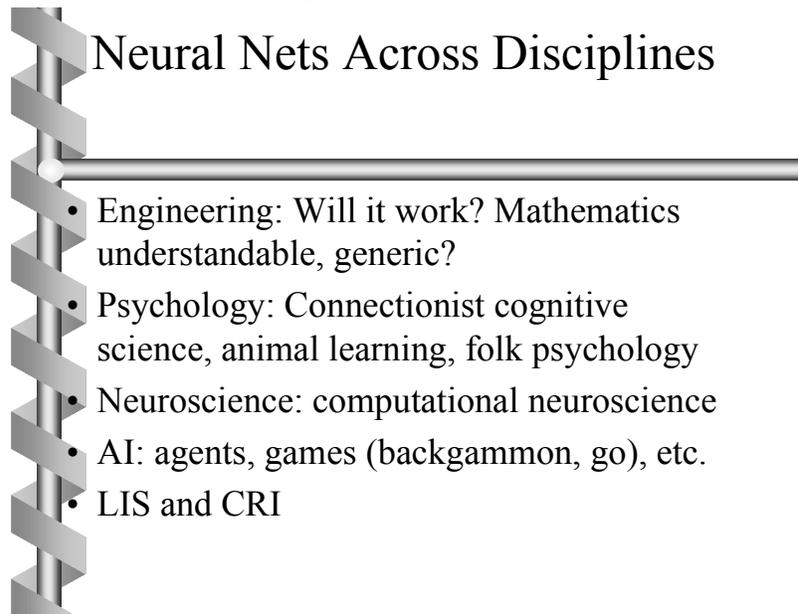

## Neural Nets Across Disciplines

- Engineering: Will it work? Mathematics understandable, generic?
- Psychology: Connectionist cognitive science, animal learning, folk psychology
- Neuroscience: computational neuroscience
- AI: agents, games (backgammon, go), etc.
- LIS and CRI

How can we do it? Well of course, the brain is made up of neural networks. And there are many neural network models already in use. We have heard about many of them here.

What is very scary is that the three communities using neural net models do not talk to each other as much as they should. The research is very fragmented today. There are people in neuroscience who have computational neuroscience models, which are designed to represent known neural circuits. There are people in psychology who have connectionist cognitive science models. And there are people in engineering who build artificial neural nets, where all they care about is "Does it work?"

These people find it hard to understand each other. I have seen Bernie Widrow and Steve Grossberg scream at each other, because they do not really appreciate each other's work... because they have different criteria for what is real work and what is bullshit. They look at the other person's work and they think that it is bullshit, because they are using a different criterion for what is good work. So Steve Grossberg is mainly asking these questions -- "Does it fit the biological circuit? Does it explain some psychological behavior?" (Grossberg is a powerful advocate of neural network research which unifies various disciplines, but these two tests have been the main drivers of his work.) The engineers, by contrast are asking "Does it work? Why does it work? What are the engineering principles involved? Does it really optimize performance?" Engineers have learned how necessary derivative calculations are to high-level general-purpose functionality; these calculations, in turn, require some use of backpropagation as part of the larger neural net designs.

Now -- to really understand the brain, intelligence in the mammal brain -- I think we must combine *all three* validation criteria. A valid model of intelligence in the brain must fit the biological data -- though it doesn't have to explain every last synapse; however, it must also fit with what we know of psychology; and it also must *work*, because the brain is a *working system* -- a highly effective, functional system. It must meet all three criteria together. So because of this idea, I helped NSF set up a new initiative a few years ago, which would allow people to get funding for this kind of cross-cutting work (among other things). It funded $20 million per year until 1999 and was called Learning and Intelligent Systems (LIS). From my point of view, the idea was to fund research to *combine* these different criteria together -- but one criterion is functionality. Where I work, in the engineering directorate, we try to build things which work.

Now since these communities do not talk to each other, some of you might not know who I am. In the engineering community, at least in the United States, almost everyone thinks of me as "He is that backpropagation person. He is that person who developed an algorithm called backpropagation back in 1974" (in my Harvard PhD thesis). That thesis is reprinted in its entirety in Werbos (1994a). Backpropagation is now used in 80% or more of the *working* applications of artificial neural nets. There are many artificial neural nets used in academia, in research papers, but for things that actually work, that are functional, solving real-world problems, 80% are based on backpropagation. You should be warned, however, that many of the popularized treatments of backpropagation oversimplify the method, and do not convey how powerful , general and flexible it really is. Until recently, the need for backpropagation in engineering designs was a major reason for the disconnect with biology; there were no proven biological mechanisms to explain how the brain itself might perform any form of backpropagation. But recent biological research has begun to fill in that particular gap. (See Bliss et al on reverse NMDA synapse, Spruston on membrane backflows, etc.)

Werbos (1994a) also begins with a chapter on why I think we are ready for a Newton-like revolution in the science of consciousness. The time has come. We have a new kind of derivative. We have new mathematics. We have new connections with Karl Pribram's kind of work. (See Werbos 1994a, 1996, 1998b). We are ready now. And in this book I talk about that. Also, for those people interested in Taoism and Buddhism, in chapter 10, I discuss the connection with those ways of thinking.

So now let me get back to the bigger question: If we want to understand the mammal brain in functional terms, first we must say what is the function. If it is not just associative memory, what is the mammal brain doing, in functional terms?

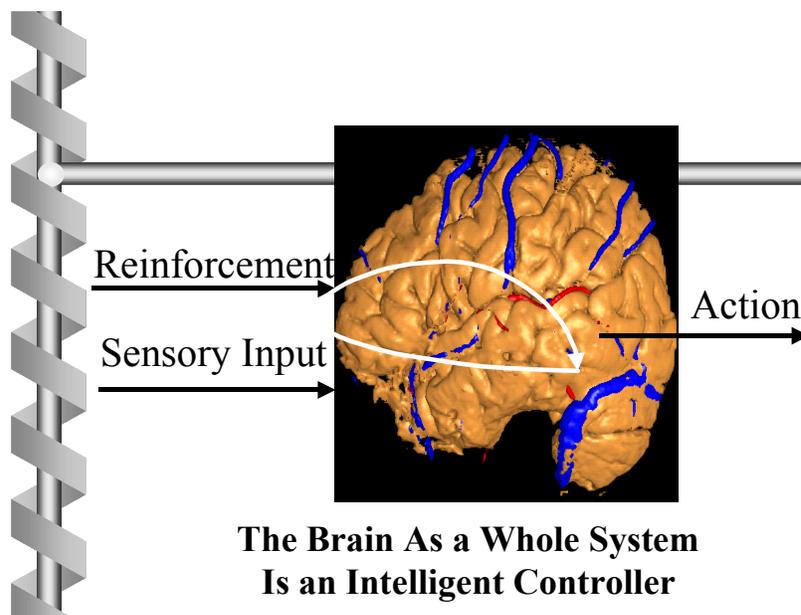

**The Brain As a Whole System
Is an Intelligent Controller**

On this next slide, I am simply saying that the brain as a whole system is what we call an intelligent controller, in engineering. The purpose of the *whole system* -- the purpose of any computing system -- is to calculate its outputs. The outputs of the brain are *actions* – what biologists call "squeezing and squirting." That is the purpose of this system, in the physical brain. And so we need to develop a mathematics of intelligent control by neural networks. Notice that I am not talking about "control of the brain;" I am talking about how the brain generates what engineers call "control signals," the signals which come out of the brain and decide on the level of "squeezing and squirting."

Let me say one other thing. Once I heard a mystic who said "You guys are all crazy. You must learn to appreciate your true self. Your true self," he said, "is much bigger than the brain and the body." And then I asked, "Well, then, what is the brain?" He said, "The brain -- it has its role -- all it is is a low level system, just to control the muscles and the glands of the body." I said, "OK, I can live with that." The brain is a controller. Now let us try to understand how such a controller can work.

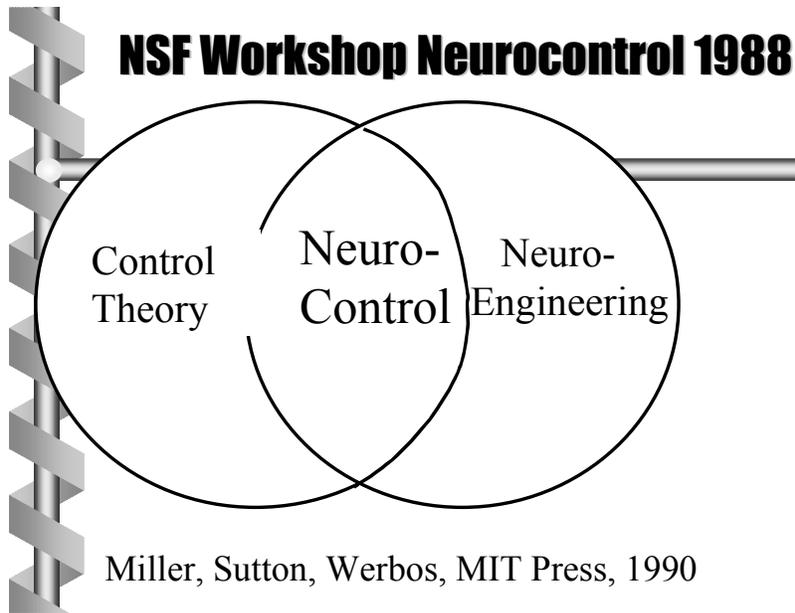

# NSF Workshop Neurocontrol 1988

Control Theory — Neuro-Control — Neuro-Engineering

Miller, Sutton, Werbos, MIT Press, 1990

When I took over the NSF program in neuroengineering 10 years ago, immediately I asked: "What do we know about neural networks for control? " We tried to survey all the ideas. We held a workshop in 1988 in New Hampshire on neurocontrol. And I invented this new word "neurocontrol." (More precisely: Allon Guez, an engineer from Drexel, coined this term for an unpublished small IEEE tutorial, and I adapted it for this use. Since then, unfortunately, some folks in biology have used the term "neural control" for a variety of different pursuits which do not even include engineering functionality.)

      In this workshop, we brought together real control theorists from engineering who know the mathematics of control -- how to make control systems that work. Brain systems are not *general* complex systems. They are a special type of system designed by nature to *work*. And so we need to use the mathematics of control systems that work. That is a very special mathematics. But we also need to know about neural networks. At this workshop we had psychologists and neuroscientists and Grossberg people. And one thing we found out: most of the neural network models out there have no hope to approximate the kind of power we see in the mammal brain. For example, there were many control models based on some old ideas from David Marr about the cerebellum. There were many models based on the idea of learning a mapping from sensory coordinates to motor coordinates. Those kind of biological models are exactly like some simple models from control theory, a class of models which are very well understood -- they work very well for certain simple problems -- but experiments have proven that even the lower level of human motor control is much more powerful than any system like that. (See the discussions of direct inverse control in Miller (et al) 1990 by myself, Jordan and Kawato. See also Werbos (1996:273-274; 1994b:698; 1999b:360:361).)

      And so in this workshop, we created this new field of neurocontrol as defined here. This slide gives a definition of neurocontrol, this word I made up. It is the subset of control theory and neural nets. We started that field.

      In this workshop, we found that there is only one class of neural network design, from engineering or psychology or biology or anywhere else, which has a hope of capturing the kind of intelligence we see in the mammal brain. This is a class of designs which some people call "reinforcement learning systems" (RLS), illustrated on the next slide.

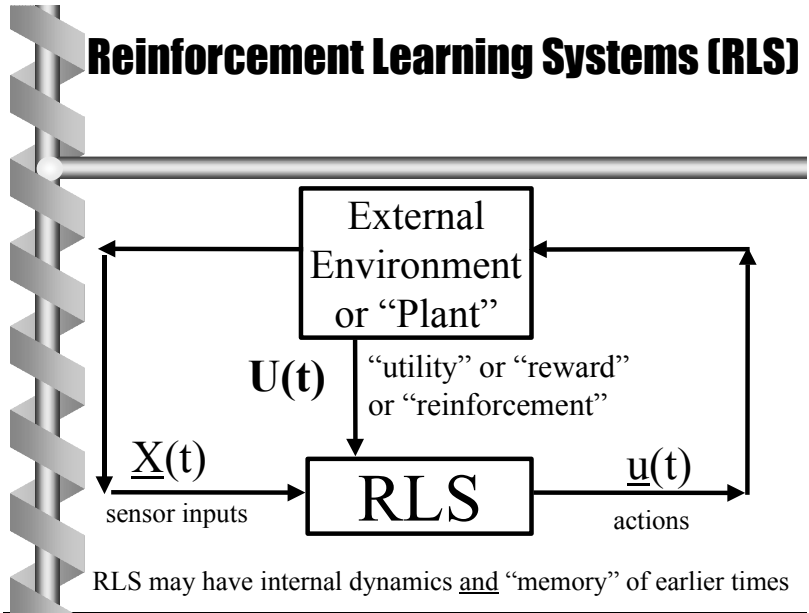

RLS may have internal dynamics and "memory" of earlier times

If you are a psychologist, this phrase "reinforcement learning" will instantly remind you of many bad old things. So I have to warn you ... I am not speaking about Skinner-type reinforcement. Also, the idea shown in this slide is somewhat simplified. This is a good starting point, but we have modified the model to account for more complicated ideas from biology and engineering. But I do not have much time to give you the complicated part today; I have to give you the simple starting point.

The idea in reinforcement learning systems (RLS) is to design an intelligent controller. Any RLS has sensor inputs. It has action or control outputs. It receives a signal of "utility" or reward. This is like pain or pleasure, perhaps. The goal is to build a system which can learn to maximize this reward signal over time. So my claim is: the mammal brain is *like* -- something like -- a reinforcement learning system.

And now I must say something very important. The mind is not *only* the intelligence. The intelligence is trying to maximize this signal (U), but this signal is not trivial. Yes, it includes pleasure and pain, but it also includes what Dr. Matsumoto was talking about -- "linkage drives" , imprinting, some kind of deep affect. The system here is actually very complex. It's also an important part of the mind. But I do not have time to talk about it today. Instead, I will give you a commercial. Karl Pribram's edited book, Brain and Values, talks a lot about this part of the mind. (See Werbos 1998b and other chapters in the same book.) Today I will only talk about the intelligence part.

If we imagine that the brain is an RLS, or something like an RLS, what does that tell us about its design? How can RLS systems actually be designed and understood, in functional mathematical terms? The slide above describes a starting point for answering these questions.

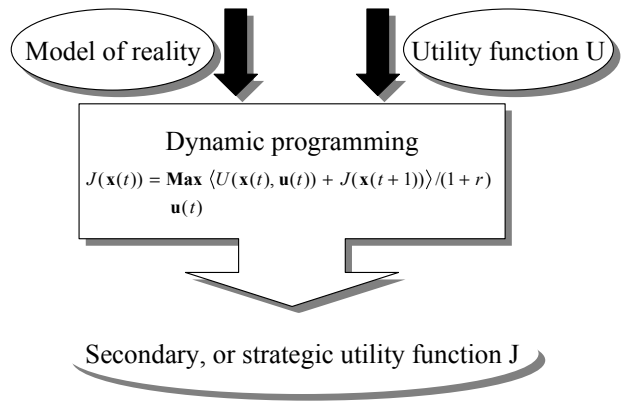

In 1968, I published an article in the journal Cybernetica (Werbos 1968), arguing that we could build reinforcement learning systems by approximating a method in control theory called dynamic programming. The brain cannot use exact dynamic programming; it is too complex for the brain. It would take a brain larger than the size of the universe to use dynamic programming to solve most everyday problems. But the idea behind the method is very interesting. In dynamic programming, we input this utility function U, and we solve for another function called J. After that, you maximize J in the short term. So U would correspond to things like pain and pleasure; J would correspond to things like learned hopes and fears. So if we build a machine based on this principle, we are building a machine that has one component which learns hopes and fears, and another part which responds to hopes and fears. With all due respect to David Chalmers, I do not think it is a "hard problem" to see the connection between this kind of design and our subjective experience. The hard problem is to make this kind of design work, and work out the details. (Note: we do have many working systems now based on these principles, but we have only just begun the resulting paradigm shift in engineering. See Werbos 1999b,c.)

Now actually, there are many, many levels of design for reinforcement learning systems. There is a whole staircase of general-purpose designs, of ever greater complexity and capability. I really do not have time to explain them all now. There is one class of design I developed back in 1971 now called "Model-Based Adaptive Critic" (MBAC). There is a new level I developed just in 1998, based on listening to Karl Pribram and changing my model, to account for the things I felt were missing after I talked to Karl. And this is still only the mammal brain. Beyond that I have some ideas - theoretical ideas -- not mathematics --

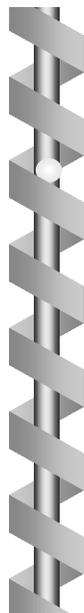
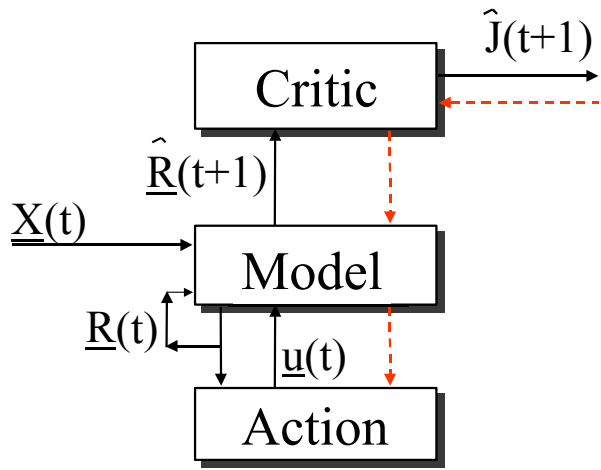

about what lies beyond.

The ideas for the Model-Based Adaptive Critic were described in great detail in a book called The Handbook of Intelligent Control (White and Sofge 1992) and there were some applications that have been developed. In the last five years, we have discovered that these are very powerful systems. For example, the design shown in the slide above is a system I proposed in 1972, in my Harvard PhD thesis proposal. This design was based on trying to translate Freud's ideas about "psychic energy" and learning into mathematics -- and that's where backpropagation really came from. The story of this is in Werbos (1994a), with some additional details in Anderson and Rosenfeld. We have recently found out that a new version of this design gives us a form of adaptive control more stable than anything else which exists now in adaptive control theory, in the linear case. (Werbos 1999c).

Even this old design from 1972 meets certain tests for a brain-like intelligent system, shown on the next slide. Five years ago, that old design was the only model of neural networks which anyone had ever implemented which meets all four tests shown here. It has an emotional or value system, a test which Dr. Matsumoto has emphasized. An intelligent control system is not a brain-like system if it does not have a

# 4 Tests For 1st-Order Model of Intelligence In the Brain

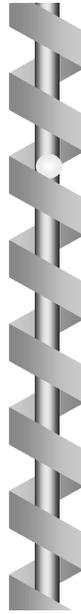

- An "Emotional" System (Values)
- An "Expectations" System (SysID)
- An Action/Motor System
- ENGINEERING FUNCTIONALITY

value system! It also had a prediction or expectation system, which Dr. Matsumoto has also talked about. And it had engineering functionality – as a general-purpose learning system. It was the first model to meet all four standards.

This is not just theory! The McDonnell-Douglas people applied an early version of this to solve a problem called making high-quality carbon-carbon composite parts. (White and Sofge 1992). These composite parts are half the cost of modern aircraft. People spend billions of dollars making these parts like cookies. PhDs baking cookies in an oven, and burning most of them. It's very expensive. McDonnell-Douglas developed a new continuous production process, but they could not control that process well enough with classical control theory, ordinary neural nets, or anything else -- but these adaptive critics were able to solve this problem, and now they can produce continuous parts. This was a big breakthrough. (Not long after that, however, White and Sofge, who developed that work at McDonnell-Douglas, moved to MIT; Boeing acquired McDonnell, and White found greater funding in the semiconductor area.)

There are many other applications I don't have time to discuss, in aerospace, in the automotive sector... Ford Motor Company has said (Business Week, Sept. 1998) that by the year 2001, every Ford car will have a neural network controller to meet air quality standards, using some algorithms that I developed... so these are working systems; it is not all theory.

Let me finish up with some citations. For the mammal level of intelligence, Karl Pribram's books -- I have some papers in there. There is also a book called Dealing with Complexity (cited in Werbos 1998b) where I discuss a new "three-brain model" based on conversations with Karl. For practical engineering applications, there are some web sites (Werbos 1999c), which include a free long paper on stability theory from the viewpoint of classical control theory. There is a paper on applications (Werbos 1999b). And then there are some papers on consciousness and on quantum theory.

To go beyond the MBAC type of model I talked about before, and to account for new things I have learned from Karl, the new model has certain characteristics. It involves neural networks which input from a field or physical networks or grids rather than just vectors. (Patent pending). It includes ways to organize a hierarchical decision system, based on a new generalization of Bellman's equation in dynamic programming. Dr. Matsumoto talked about a hierarchical system here today. Karl Pribram discussed this in his book with Miller and Galanter on Tasks and the organization of behavior. Now there is a mathematical implementation of Karl's ideas, and a new form of dynamic programming to implement these ideas in a learning system. We also have some things called imagination networks... there are many new things I cannot show you for reasons of time.

## 2.3. Additional Comments on Quantum Theory and the Mind

Now: two slides on quantum theory, and some comments on mind and reality.

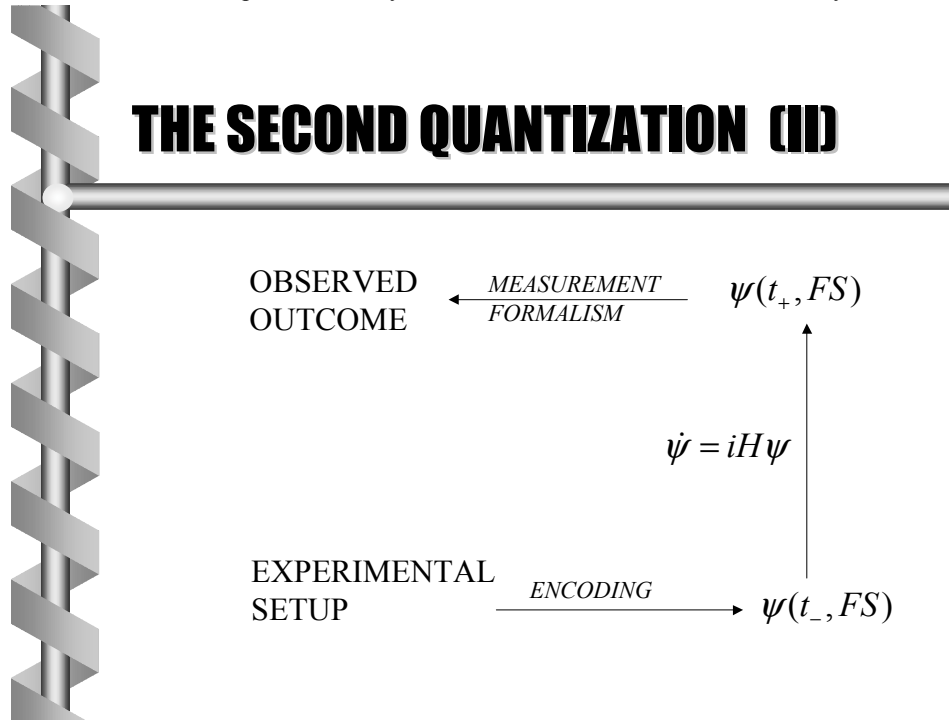

This slide depicts quantum field theory, in the second quantization. This is the quantum field theory which most people work with today. They use a wave function, which is a function of a very complex space called Fock space. There is a kind of "Schrodinger equation" - not the old Schrodinger equation -- which evolves over time, and there is a measurement formalism. The standard ideology, the standard form of quantum mechanics, says that you need a conscious observer, a metaphysical observer, but there were new experiments done by Mandel and Ou, reported in Scientific American in 1992 (June) which showed that you can get measurement effects without a conscious observer. So there is empirical evidence that we need a quantum theory without observers. This is experiment -- this is not philosophy. Where can we get such a quantum theory?

I cannot explain quantum theory in one minute! But I can give some citations. Werbos (1998a: section 6) includes three alternatives to the usual formulation of the functional integral approach – one a slight reformulation of Schwinger's ideas, to make them more compact and parsimonious, but another one very crazy and heretical – providing a more formal basis for revisiting the possibility of realism, drawing on some of the old ideas of Einstein and DeBroglie. Werbos (1999b) provides some of the conceptual background; section 6 of that paper also talks about the three alternatives, and possible testable implications. (Section 3 of this paper will add a new idea on those lines.) For example, there is a possibility that quarks could be bosons... there is a way you could do it. It sounds crazy. But I think I know how.

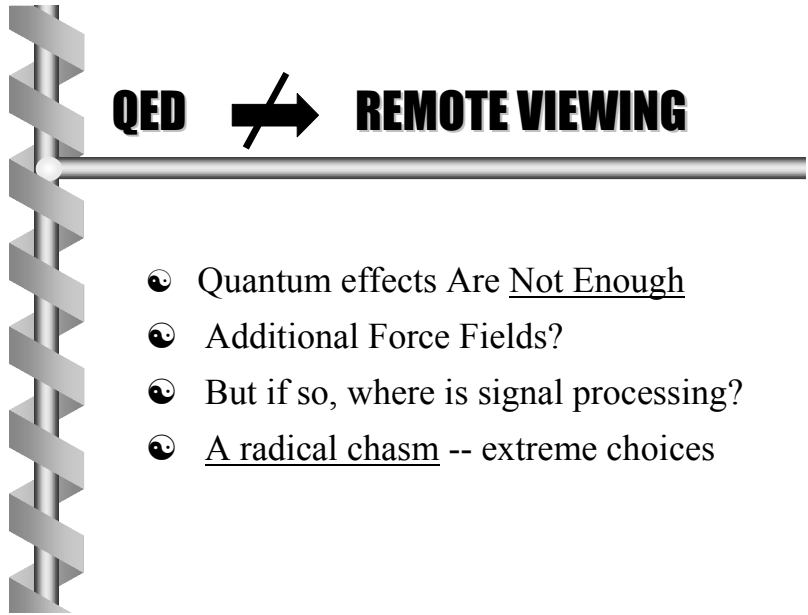

- Quantum effects Are <u>Not Enough</u>
- Additional Force Fields?
- But if so, where is signal processing?
- <u>A radical chasm</u> -- extreme choices

One last slide. Many people at this conference have expressed hope that quantum mechanics might explain things like remote viewing or like the collective unconscious of Jung -- wild, crazy things. I would like to point out that no form of quantum mechanics can explain something like remote viewing. It doesn't matter whether you take Bohmian or my kind or Schwinger's kind or Copenhagen... because all these different forms of quantum mechanics produce about the same quantum electrodynamics ... they yield the same predictions, essentially, for the case of quantum electrodynamics (QED). If you consider electrodynamics, that is not enough to generate remote viewing. We know what is possible with QED. The world has spent billions of dollars trying to use QED in the military to see things far away. We cannot do it. So if you want to explain strange things like remote viewing, the only way is by assuming strange force fields and strange signal processing . You have a choice. There is a great chasm. It is a binary choice. You cannot do it a fuzzy way. Either you give up on these phenomena -- you give up on all that stuff -- or else you have to open yourself up to *really* crazy things, much more than just quantum theory. Crazy things like letting me stay here... and I thank Kunio for allowing such a crazy thing.

## 3. Recent Extensions

This section will *not* give more detailed explanations of the ideas discussed above; see the references for such explanations. Instead, it will give a condensed summary of some new thinking, stimulated by discussions at this conference and at the Quantum Mind conference in Arizona.

### 3.1. Comments on Consciousness Qua Wakefulness or Awareness

Because wakefulness and awareness are major aspects of brain functioning they are, of course, addressed in the models I mentioned above.
       In one of Pribram's recent conferences, there was a debate between Pribram, McClelland, Alkon and  myself on the functional significance of sleep states. From my earliest papers, I agreed with LaBerge that dreams provide a *simulation* capability, essential to the training of any imaginative intelligent controller. Working RLS systems have demonstrated this kind of capability. Additional states are required to facilitate memory consolidation or generalization from memory – a topic related to what is called "memory-based learning" or "syncretism" on the engineering side; McClelland has argued that this involves a transfer from hippocampus to neocortex during dreams, but Karl and I argued that it may instead involve a harmonization between different types of cell *within* these two structures, during other kinds of sleep states. A key technical point is that local and global representations both exist within both organs. Furthermore, dreams and the hippocampus have long been known to have other functions beyond this hypothesized memory function.

Regarding awareness and attention – I thank Bernie Baars for drawing my attention to some of the recent literature by authors like himself and Legothetes, which I need to study further. Attention is clearly much more than a matter of importance weighting or "salience," as in the older models. In my view, it is the key mechanism for "labeling" the variables monitored by major fields in the neocortex; for an example of how important this might be, see the paper by Olhausen and Koch in Arbib. More precisely, this kind of object "labeling" is the kind of machinery needed to use multiplexing to implement the "ObjectNet" design (patent pending) discussed in Werbos 1999c. Any efficient multiplexing system results in synchronized "object binding," without any need for reverberatory attractors and other such mechanisms popular in neuroscience today; the challenge for design (or functional understanding) is not with the binding per se, but the management and choice of what is bound *to*. Current evidence (see papers in Arbib) suggests that the pulvinar plays a crucial role in this function.

### 3.2. Discussions at the Arizona Conference

I am very grateful to Stuart Hameroff and the Arizona group for inviting me to speak at that conference, despite my known skepticism about ORCH as such.

At Arizona, I argued that true quantum computing effects probably are not relevant to a functional understanding of the brain. This does not mean that quantum mechanics as such is irrelevant. Quantum mechanics is important to understanding how molecules work, just as it is important to understanding how quantum dots and Josephson junctions can be used to implement classical NOT gates and AND gates, etc. But we would call that "quantum devices," not "quantum computing," in modern terminology. If a computer is based on quantum devices and ordinary field effects (such as those Pribram has often discussed), this is still quite consistent with the class of quasi-Turing-machine model we are now working with to understand the mammal brain level of intelligence. But for true quantum computing, as now defined, there must be some exploitation of *coherence* or *quantum entanglement* effects to serve a systems-level computational purpose. Many people have already talked about the difficult, unproven physics of trying to imagine how brains could create and maintain quantum entanglement, but very little attention has gone into the even more serious issue of trying to imagine what kind of computational purpose might be served by such a system in the brain.

As an honest skeptic, perhaps my first duty is to issue a challenge to the quantum brain believers – to give an example of what they *might* try to prove, to overcome my skepticism. From all I have read and thought about, I can only imagine two ways that a "quantum computing" capability in the brain might really affect general-purpose intelligence. One would be the evolution of a "quantum associative memory" neuron. Could one really train a single neuron to learn simple functions like XOR or Minsky's old parity mapping challenge? These are not "natural" problems – but if an individual neuron really had the ability to use molecular quantum computing to achieve associative memory, it should have the ability to learn such relations. If it does not... then what are the hypothesized quantum effects within the cell doing anyway? A second possibility would be that of a "superfast recurrent network" (SFRN), an alternative approach to quantum computing (a form of quantum neural network) proposed in Werbos (1997); however, that hypothetical possibility has yet to be fully understood in engineering terms, let alone mapped into biology.

Crucial to the idea of an SFRN is the old insight, originally due to myself (Werbos 1973, 1989) and DeBeauregard, that the paradoxes of quantum theory can be understood as the result of causality running backwards through time at the microscopic, quantum level. (This is similar in spirit to Cramer's later "transactional interpretation," but Cramer invokes nonlocality, which is unnecessary here.) Penrose cited us both in *Shadows of the Mind*, and Hameroff showed a slide from Penrose conveying the idea very vividly. Various people went on to argue that new evidence (from Libet, Radin and Bierman) shows that the brain can respond ¼ of a second *before* a stimulus, and that something like an SFRN might be present in the brain. Parts of this evidence were surprisingly convincing to me, personally, and they posed more acutely the need to revisit the concept of SFRN and backwards causality. Mari Jibu also pointedly challenged us to explain more precisely how we think the interface actually works between "microscopic" time symmetry and the macroscopic arrow of time.

As a caveat, Josephson reminded people that my negative comments pertain only to the brain – not to the "soul," a subject of great interest to many but beyond the scope of the present discussion.

### 3.3. Revisions of My Views of Quantum Effects

Word limits here require that I must assume the reader has full knowledge of the references. The views here are not only personal but highly tentative.

Many issues which *seem* real, in debates on quantum theory, disappear when one considers recent experiments. (In addition to the quantum computing work mentioned above, Y. Shih and K. Alley of Maryland have important results.) For example, one may worry about what happens after two measurements, A(t) and B(s), at times s and t, at the discontinuity where s=t. But real measurements take some time; when one approaches such a discontinuity, one predicts the result simply by representing the polarizers or whatever in a more complete fashion, as potentials or particles, affecting the Schrodinger equation, and chucking out the metaphysical observer formalism. This is like the original Von Neuman-Wigner approach discussed by Stapp at Arizona. This is what actually works. As a practical matter, one always expects to get the right result if one applies the measurement and setup formalisms only to the ultimate, asymptotic, commutative inputs and outputs of an experiment; the measurement formalism may *sometimes* work in describing what happens in the middle, but there is no general guarantee.

Both the functional integral approach, and the variations which I have proposed (Werbos 1998a,1999a), assume an underlying symmetry in time at the microscopic level (leaving aside the superweak interactions). In answer to Mari Jibu's question, I would argue that all the usual experiments in quantum theory can be reduced to something I call "the standard paradigm." In this paradigm, everything is ultimately reduced to a scattering experiment. The inputs are represented by some set of measurement operators, *and by* the actual values of the corresponding variables. (For example, the experimenter may control the momentum of every incoming particle.) The outputs are represented by another set of measurement operators, but the experimenter cannot *decide* the values of those variables; he may only observe them. Thus there is a clear-cut asymmetry between the input situation and the output situation. In practice (in my *definition* of "standard paradigm"), the outgoing measurement operators all commute with each other; in fact, they are really nothing but particle counters, which measure particles with energy $E \gg kT$, where T is the temperature of the counter. (Polarizers and such may be considered as internal parts of the experiment.)

The functional integral approaches and the second quantization essentially agree completely, for experiments which can be reduced to the standard paradigm. We cannot do the usual Bell's Theorem experiments in reverse time, because these counters do not emit energetic particles in reverse time. Why not, if physics is symmetric in time at the microscopic level? Actually, this question is mathematically almost equivalent to the classical question about what happens to a rock on the floor. Why do we not see rocks flying up from the floor, following a time-reversed movie of how they fall to the floor? The answer is simple: there is only a tiny probability that the atoms under the rock will happen to move in the same direction (up) and push the rock up. For similar reasons, it is rare that an $E \gg kT$ counter would emit a particle in reverse time. The puzzling thing is that we ever see such an event in forwards time; this otherwise improbable thing is due to the experimenter exploiting the availability of time-forwards free energy, which ultimately comes from sunlight pouring down on earth – a macroscopic boundary condition.

For experiments within the scope of the standard paradigm, backwards time communication of macroscopic information is impossible. It is impossible, in part, because it would allow a violation of Eberhart's Theorem on the impossibility of communicating information faster than the speed of light (FTL). Eberhart's Theorem does not depend on conventional wisdoms about causality and such; it only depends on the basic assumption that equal time commutators are zero. The *concept* of backwards causality and equilibration across space-time may provide a useful *understanding* of what is possible with quantum computing within the standard paradigm; thus it may still permit development of some kind of useful SFRN, as a way of speeding up certain very general computations. However, such designs could all be reformulated (albeit awkwardly) within the usual formalisms of traditional quantum computing, rooted in the second quantization. There is no possibility of communicating macroscopic information back through time.

There are two possible loopholes here which merit further thought. First, what about "stochastic infrared quantum computing?" What if one output channel ends in a controlled polarizer, follower by an $E<kT$ or $E=kT$ "counter"? Many experiments were done by Planck and Einstein in that regime, but perhaps a modern analysis might be interesting. Second, Eberhart's Theorem implicitly assumes that information is represented by bare operators, not renormalized physical operators. Gerhard Hegerfeldt, of the University of Gottingen, has shown that FTL communication can actually occur, in limited circumstances, due to renormalization effects. What if the renormalization were more drastic, as with wave functions of electrons in superconductors, which are extended very far in space?
(Final two paragraphs not attached.)